# The onset of star formation 250 million years after the Big Bang

Takuya Hashimoto*,1,2, Nicolas Laporte3,4, Ken Mawatari1, Richard S. Ellis3, Akio. K. Inoue1, Erik Zackrisson5, Guido Roberts-Borsani3, Wei Zheng6, Yoichi Tamura7, Franz E. Bauer8,9,10, Thomas Fletcher3, Yuichi Harikane11,12, Bunyo Hatsukade13, Natsuki H. Hayatsu12,14, Yuichi Matsuda2,15, Hiroshi Matsuo2,15, Takashi Okamoto16, Masami Ouchi11,17, Roser Pelló4, Claes-Erik Rydberg18, Ikkoh Shimizu19, Yoshiaki Taniguchi20, Hideki Umehata13,20,21, Naoki Yoshida12,17

[1]Department of Environmental Science and Technology, Faculty of Design Technology, Osaka Sangyo University, 3-1-1, Nagaito, Daito, Osaka 574-8530, Japan

[2]National Astronomical Observatory of Japan, 2-21-1 Osawa, Mitaka, Tokyo 181-8588, Japan

[3]Department of Physics & Astronomy, University College London, Gower Street, London WC1E 6BT, UK

[4]IRAP, Université de Toulouse, CNRS, UPS, CNES, 14, avenue Edouard Belin, F-31400 Toulouse, France

[5]Department of Physics and Astronomy, Uppsala University, Box 516, SE-751 20 Uppsala, Sweden

[6]Department of Physics & Astronomy, Johns Hopkins University, 3400 N Charles St, Baltimore, MD 21218 USA

[7]Division of Particle and Astrophysical Science, Graduate School of Science, Nagoya University, Furo-cho, Chikusa-ku, Nagoya 464-8602,Japan

[8]Instituto de Astrofísica and Centro de Astroingeniería, Facultad de Física, Pontificia Universidad

Católica de Chile, Casilla 306, Santiago 22, Chile

[9]Millennium Institute of Astrophysics (MAS), Nuncio Monseñor Sótero Sanz 100, Providencia, Santiago, Chile

[10]Space Science Institute, 4750 Walnut Street, Suite 205, Boulder, Colorado 80301

[11]Institute for Cosmic Ray Research, The University of Tokyo, 5-1-5 Kashiwanoha, Kashiwa, Chiba 277-8582, Japan

[12]Department of Physics, Graduate School of Science, The University of Tokyo, 7-3-1 Hongo, Bunkyo, Tokyo, 113-0033, Japan

[13]Institute of Astronomy, The University of Tokyo, 2-21-1 Osawa, Mitaka, Tokyo 181-0015, Japan

[14]European Southern Observatory, Karl-Schwarzschild-Str. 2, 85748 Garching bei München, Germany

[15]Department of Astronomical Science, School of Physical Sciences, The Graduate University for Advanced Studies (SOKENDAI), 2-21-1, Osawa, Mitaka, Tokyo 181-8588, Japan

[16]Department of Cosmosciences, Graduates School of Science, Hokakido University, N10 W8, Kitaku, Sapporo 060-0810, Japan

[17]Kavli Institute for the Physics and Mathematics of the Universe (WPI), Todai Institutes for Advanced Study, The University of Tokyo, 5-1-5 Kashiwanoha, Kashiwa, Chiba 277-8583, Japan

[18]Universität Heidelberg, Zentrum für Astronomie, Institut für Theoretische Astrophysik, Albert-Ueberly-Str. 2, D-69120 Heidelberg, Germany

[19]Theoretical Astrophysics, Department of Earth & Space Science, Osaka University, 1-1 Machikaneyama, Toyonaka, Osaka 560-0043, Japan

[20] *The Open University of Japan, 2-11 Wakaba, Mihama-ku, Chiba 261-8586, Japan*

[21] *The Institute of Physical and Chemical Research (RIKEN), 2-1 Hirosawa, Wako-shi, Saitama 351-0198, Japan*

**A fundamental quest of modern astronomy is to locate the earliest galaxies and study how they influenced the intergalactic medium a few hundred million years after the Big Bang[1–3]. The abundance of star-forming galaxies is known to decline[4,5] from redshifts of about 6 to 10, but a key question is the extent of star formation at even earlier times, corresponding to the period when the first galaxies might have emerged. Here we present spectroscopic observations of MACS1149-JD1[6], a gravitationally lensed galaxy observed when the Universe was less than four per cent of its present age. We detect an emission line of doubly ionized oxygen at a redshift of 9.1096±0.0006, with an uncertainty of one standard deviation. This precisely determined redshift indicates that the red rest-frame optical colour arises from a dominant stellar component that formed about 250 million years after the Big Bang, corresponding to a redshift of about 15. Our results indicate the it may be possible to detect such early episodes of star formation in similar galaxies with future telescopes.**

Between March 2016 and April 2017 we performed observations of MACS1149-JD1 at the Atacama Large Millimeter/submillimeter Array (ALMA), targeting the far-infrared oxygen line, [OIII] 88 $\mu$m, and dust continuum emission over a broad wavelength range consistent with its photometric redshift range ($z = 9.0 - 9.8$)[4,7–9]. The [OIII] line is clearly detected at a redshift $z = 9.1096 \pm 0.0006$ with a peak intensity $129.8 \pm 17.5$ mJy km s$^{-1}$ beam$^{-1}$ corresponding to the significance level of $7.4\sigma$ (Figure 1, Table 1). The spatial location is coincident with the rest-frame ultraviolet (UV) continuum emission detected by *HST* indicating that the [OIII] line arises from a star forming galaxy. The line has a luminosity of $(7.4 \pm 1.6) \times 10^7 \times (10/\mu)$ $L_\odot$, where, for the lensing magnification, we adopt a fiducial value of $\mu = 10$ (See Methods). Its full width at half maximum (FWHM) is $154 \pm 39$ km s$^{-1}$ representative of that seen in other high $z$ galaxies[10]. The [OIII] emission is spatially resolved, and its deconvolved size is $(0''.82 \pm 0''.25) \times (0''.30 \pm 0''.14)$. The corresponding intrinsic size is $(3.7 \pm 1.1)/\sqrt{\mu}$ [kpc] $\times$ $(1.4 \pm 0.9)/\sqrt{\mu}$ [kpc] under the assumption that lensing effects are equal for the major and minor axes. Assuming that MACS1149-JD1 is a dispersion-dominated system, we derive the dynamical mass of $(4 \pm 3) \times (10/\mu)^{0.5} \times 10^9$ $M_\odot$.

No dust continuum is detected above a $3\sigma$ upper limit of $S_{\nu,90\mu m} < 5.3 \times (10/\mu)$ $\mu$Jy beam$^{-1}$, where $S_{\nu,90\mu m}$ corresponds to the flux density at the rest-frame wavelength of 90 $\mu$m. For a dust temperature $T_\mathrm{d} = 40$ K and emissivity index $\beta_\mathrm{d} = 1.5$, the total infrared luminosity is $< 7.7 \times 10^9 \times (10/\mu)$ $L_\odot$ after correcting the contribution of the cosmic-microwave-background (CMB)[11]. Assuming a representative dust mass absorption coefficient[12], we derive a $3\sigma$ upper limit on the dust mass of $5.3 \times 10^5 \times (10/\mu)$ $M_\odot$.

The spectral energy distribution (SED) of MACS1149-JD1 has a prominent excess signal in the *Spitzer* IRAC channel 2 band at 4.5 microns[6]. Previous spectroscopic studies of such 'IRAC-excess' sources[13–15] have claimed that such an excess likely arises from intense [OIII] 5007 Å line emission. However, the origin of the IRAC-excess of MACS1149-JD1 remained unclear due to the inaccuracy of the photometrically estimated redshift[6,7,16]. Our precisely-determined redshift of $z = 9.1$ now rules out the [OIII] 5007 Å line as the source of the excess since the line is at a wavelength with $< 1\%$ transmission in the IRAC channel 2 bandpass. Given recent evidence that [OIII] 88 $\mu$m may not be an accurate indication of the systemic velocity of distant star-forming galaxies[17], we also present more tentative evidence for Lyman-$\alpha$ (Ly$\alpha$) emission from MACS1149-JD1 using the X-shooter[18] spectrograph at the European Southern Observatory's Very Large Telescope using 6.5 hours of data taken between February and April 2017. A $4\sigma$ significant detection at $\lambda = 12,271.51$ Å , corresponding to Ly$\alpha$ at $z = 9.0942 \pm 0.0019$, provides a further valuable constraint on the source redshift.

Our detailed analysis (see Methods) shows that strong H$\beta$ + [OIII] 4959 Å lines in IRAC channel 2 cannot reproduce the excess within the likely source redshift range because, for the required young ages, the IRAC channel 1 flux would also be significantly boosted by strong nebular continuum emission and [OII] 3727 Å line emission. Such young models are further disfavoured given additional constraints from the measured UV continuum slope and the ALMA upper limit on the dust attenuation. We conclude therefore, that the IRAC excess can only arise from the stellar continuum around the Balmer break at $\sim 4000$ Å.

Despite the fact MACS1149-JD1 is elongated by the effects of gravitational lensing, there is some evidence of two components in the *HST* image (Figure 1), so it is conceivable the IRAC 4.5 micron excess originates from a separate source. The peak of the ALMA [OIII] emission is coincident with the brighter component (Figure 1), whereas the less well-located Ly$\alpha$ emission and IRAC flux could arise in either component. However, the wide wavelength coverage of the SED enables us to reject the possibility that a low redshift interloper contributes to the 4.5 micron excess (see Methods for detailed analyses).

Independently of the SED, the ALMA [OIII] emission line flux provides a valuable measure of the star formation rate (SFR) at the epoch of observation[10,19]. To reproduce the SED and the size of the Balmer break, it is necessary to consider star formation histories extending $\simeq 300$ Myr earlier, corresponding to the onset of galaxy formation at redshifts $z \simeq 15$ (Figures 2 and 3). Our fiducial model has a dominant old component whose SFR declined, thereby reproducing the Balmer break, with a much younger component which reproduces the strength of both Ly$\alpha$ and [OIII] 88 $\mu$m emission. Although the detailed parameters of these two components cannot be uniquely determined with the current data, we demonstrate in the Methods that a dominant earlier phase of star formation whose activity subsequently declined is an inescapable conclusion necessary to reproduce the Balmer break. Indeed, the bulk of the stellar mass of MACS1149-JD1 observed at $z = 9.1$ is contributed by this early component ($\simeq \times 10^9 \times (10/\mu)\ M_\odot$, Figure 3). Its onset at $z = 15.4^{+2.7}_{-2.0}$ thus represents the logical epoch of formation of MACS1149-JD1[6].

During this dominant first episode of star formation, feedback processes will likely weaken or even terminate star formation activity, leading to a quiescent phase until a second episode when gaseous inflow can rejuvenate star formation at $z = 9.1$ leading to the detection of [OIII] 88 $\mu$m emission. Intense UV radiation from this early generation of massive stars may create an ionised bubble[20–22] surrounding MACS1149-JD1 displacing the damping wing of Ly$\alpha$ absorption by intergalactic neutral hydrogen blueward by $\simeq 500$ km s$^{-1}$, facilitating transmission through the predominantly neutral IGM. Although somewhat speculative given its tentative nature, inflowing gas may then provide an explanation for the blueshift of Ly$\alpha$ with respect to [O III] 88 $\mu$m by several hundred km s$^{-1}$ [23,24] (see Methods).

MACS1149-JD1 is already a well-established galaxy. Although we are observing a secondary episode of star formation at $z = 9.1$, the galaxy formed the bulk of its stars at a much earlier epoch. Our results indicate it may be feasible to directly detect the earliest phase of galaxy formation, beyond the redshift range currently probed with *HST*, with future facilities such as the *James Webb Space Telescope*. In addition, our observations demonstrate the great power of ALMA[19] to spectroscopically identify galaxies at $z > 9$, showing that ALMA will also play a central role in investigating the first-generation galaxies.

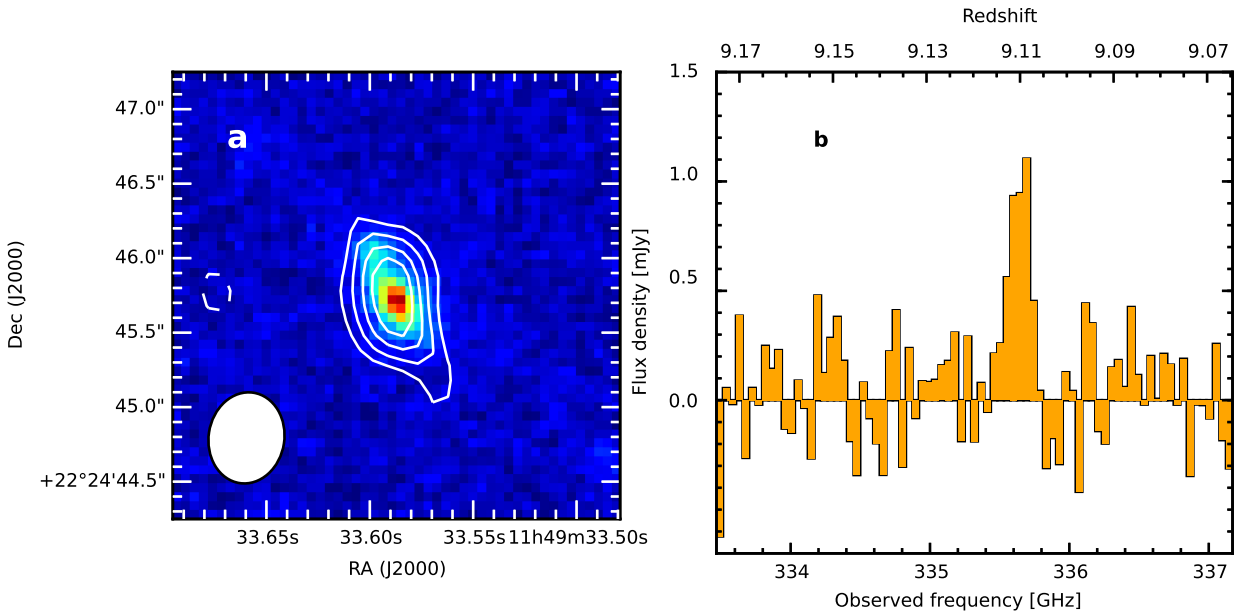

**Figure 1 | ALMA [OIII] contours and spectrum of MACS1149-JD1.** (a) Zoom on an *HST* image (F160W) with the ALMA [OIII] contours overlaid. Contours are drawn at $1\sigma$ intervals from $\pm 3$ to $+6\sigma$ where $\sigma = 17.5$ mJy km s$^{-1}$ beam$^{-1}$. Negative contours are shown by the dashed line. Ellipse at the lower left corner indicates the synthesized beam size of ALMA. (b) ALMA [OIII] 88 $\mu$m spectrum in frequency space with a resolution of $\sim 42$ km s$^{-1}$.

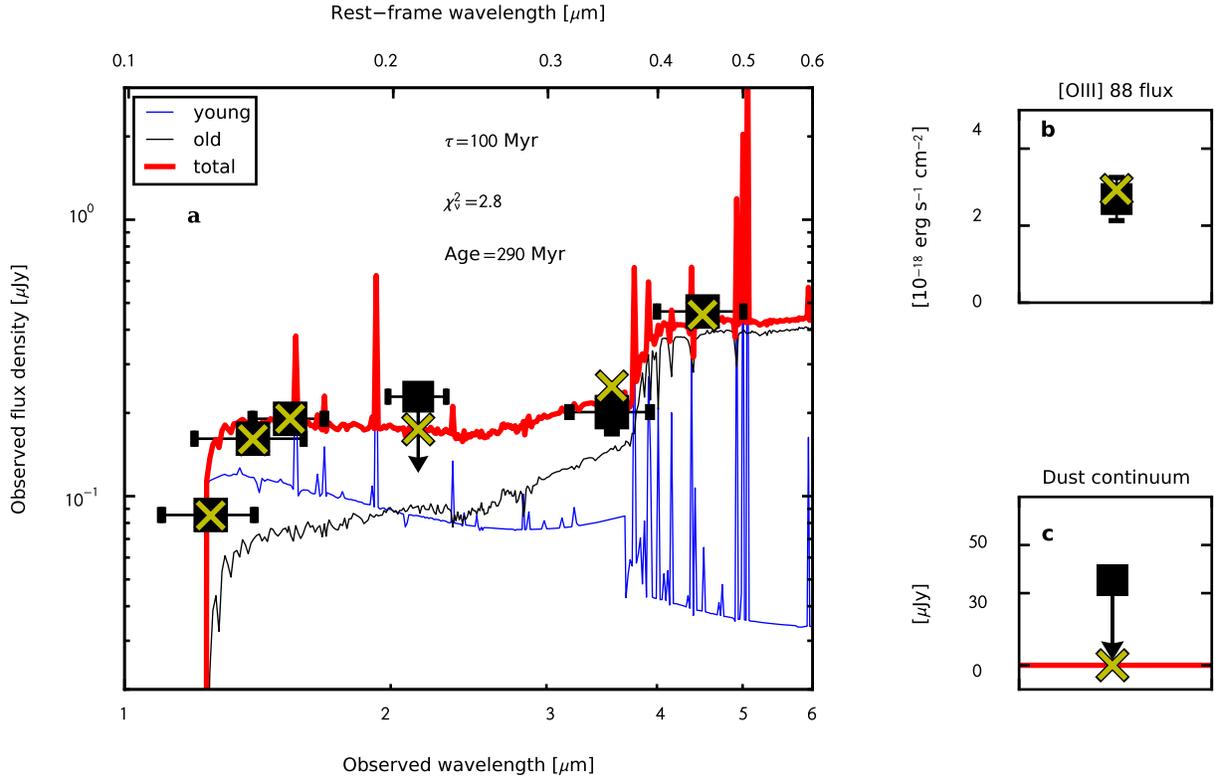

**Figure 2 | The Spectral Energy Distribution of MACS1149-JD1.** The [O III] flux provides a constraint on the star formation rate at the epoch of observation, while the Spectral Energy Distribution (SED) provides valuable evidence of its earlier star formation history. The model shown involves a burst of star formation of duration $\tau = 100$ Myr within the interval $12 \lesssim z \lesssim 15$ and reproduce the required excess flux at 4.5$\mu$m. A much younger component reproduces the strength of [O III] emission observed at $z \simeq 9$. (a) Black squares show the observations: F125W, F140W, and F160W data from *HST*[7], a $2\sigma$ upper limit for the $K_s-$band from VLT/HAWK-I[16], and 3.6$\mu$m and 4.5$\mu$m fluxes from *Spitzer*/IRAC[7]. Horizontal and vertical error bars show the wavelength range of the filters and 1$\sigma$ measurement uncertainties, respectively. The red solid line

indicates the SED model and the corresponding magnitudes are shown via yellow crosses. Blue and black lines represent the contributions from the young and old component, respectively. (b) The black square is the observed [O III] emission line flux and its $1\sigma$ uncertainty, while the yellow cross indicates the model prediction. (c) The black square shows the $2\sigma$ upper limit for the dust continuum flux density, and the yellow cross the model prediction.

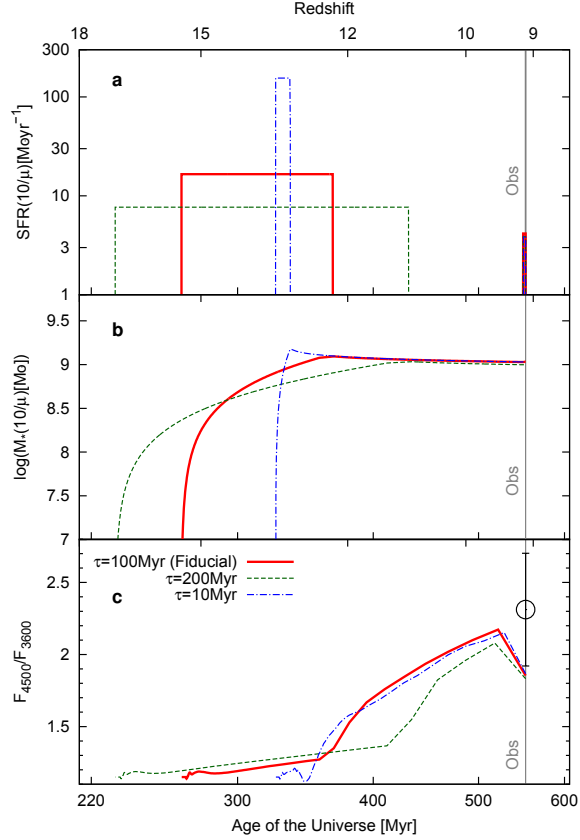

**Figure 3 | Demonstration of how a dominant phase of early star formation is necessary to reproduce the SED of MACS1149-JD1.** We show the star formation rate (a), stellar mass assembly (b) and Balmer break (c) as a function of redshift for three possible models. $F_{4500}$ and $F_{3600}$ indicate the flux densities at the rest-frame wavelengths of $4,500$ Å and $3,600$ Å, respectively. In each panel, the blue dot-dashed, green dashed, and red solid lines correspond to the models with star formation bursts of duration 10, 100, and 200 Myr, respectively. Each is capable of reproducing the Balmer break observed at $z = 9.1$ (shown by the black data point). The full SED fits shown in Figure 2 corresponds to our fiducial model corresponding to a burst of duration 100 Myr. The bulk of the stellar mass in MACS1149-JD1 observed at $z = 9.1$ was produced within a short period corresponding to the redshift interval $12< z <16$. Further details of these models and other assumptions are provided in the Methods.

**Table 1 | Properties of MACS1149-JD1.**

| Parameters | Values |
|---|---|
| RA | $11:49:33.58$ |
| Dec | $+22:24:45.7$ |
| Redshift $z_{\rm [OIII]}$ | $9.1096 \pm 0.0006$ |
| [OIII] line width FWHM$_{\rm [OIII]}$ (km s$^{-1}$) | $154 \pm 39$ |
| [OIII] luminosity $L_{\rm [OIII]}$ ($10^7$ $L_\odot$) | $(7.4 \pm 1.6) \times (10/\mu)^a$ |
| 90 $\mu$m continuum flux ($\mu$Jy beam$^{-1}$) | $< 5.3 \times (10/\mu)^a$ ($3\sigma$) |
| Star formation rate ($M_\odot$ yr$^{-1}$) | $4.2^{+0.8}_{-1.1} \times (10/\mu)^a$ |
| Stellar mass ($10^9 M_\odot$) | $1.08^{+0.53}_{-0.18} \times (10/\mu)^a$ |
| Dust mass ($10^5$ $M_\odot$) | $< 5.2 \times (10/\mu)^a$ ($3\sigma$)$^b$ |
| Dynamical mass ($10^9 M_\odot$) | $(4 \pm 3) \times (10/\mu)^{0.5}$ $^{a,c}$ |
| Redshift $z_{\rm Ly\alpha}$ | $9.0944 \pm 0.0019$ |
| Velocity offset $\Delta v_{\rm Ly\alpha}$ (km s$^{-1}$) | $-450 \pm 60$ |
| Ly$\alpha$ line width FWHM$_{\rm Ly\alpha}$ (km s$^{-1}$) | $144 \pm 56$ |
| Ly$\alpha$ luminosity $L_{\rm Ly\alpha}$ ($10^7$ $L_\odot$) | $(12.4 \pm 3.2) \times (10/\mu)^a$ |

$^a$ $\mu$ is the lensing magnification

$^b$ We assume dust temperature $T_{\rm d} = 40$ K, spectral index $\beta_{\rm d} = 1.5$, and the dust emitting region of a single beam size

$^c$ The value under the assumption that lensing effects are equal for the major and minor axes

**Acknowledgements** We thank K. Umetsu for a discussion of gravitational lensing models, K. Nakanishi, F. Egusa, and K. Saigo for discussions in handling with ALMA data, S. Kikuchihara for supporting MOSFIRE observations, H. Yajima and A. Zitrin for helpful discussions. We acknowledge support from: NAOJ ALMA Scientific Research Grant Number 2016-01A (T.H.and A.K.I.); European Research Council Advanced Grant FP7/669253 (N.L. and R.S.E.) and 339177 (C.E.R.); KAKENHI grants 26287034 and 17H01114 (K.M. and A.K.I.), 17H06130 (Y.Tamura), 17H04831 (Y.M.), 16H01085 (T.O.), 16H02166 (Y.Taniguchi), 15K17616 (B.H.), 17K14252 (H.U.), JP17H01111 (I.S.), 16J03329 (Y.H.), and 15H02064 (M.O.); the grant CONICYT-Chile Basal-CATA PFB-06/2007, FONDECYT Regular 1141218 (F.E.B.); NAOJ Visiting Fellow Program (N.H.H.). ALMA is a partnership of ESO (representing its member states), NSF (USA) and NINS (Japan), together with NRC (Canada), NSC and ASIAA (Taiwan), and KASI (Republic of Korea), in cooperation with the Republic of Chile. The Joint ALMA Observatory is operated by ESO, AUI/NRAO and NAOJ. This work incorporates with observations with ESO Telescopes at the La Silla Paranal Observatory. This work is also in part based on observations made with the *Spitzer* Space Telescope, which is operated by the Jet Propulsion Laboratory, California Institute of Technology under a contract with NASA, as well as the NASA/ESA *Hubble* Space Telescope, obtained at the Space Telescope Science Institute (STScI), which is operated by the Association of Universities for Research in Astronomy, Inc. under NASA contract NAS 5-26555.

**Author Contributions** T.H., N.L., R.S.E., and A.K.I. wrote the paper. T.H. and Y.Tamura reduced and analyzed ALMA data. T.H. produced Figs. 1 and 2 and Extended Data Figs. 1, 3-5. N.L. reduced and analyzed X-Shooter data, and produced Extended Data Figs. 2 and 7. K.M. and E.Z. performed SED fitting analyses. K.M. produced Fig. 3 and Extended Data Fig. 6. W.Z. carried out the astrometry analysis on *HST* and IRAC data. N.L. and C.E.R. performed lensing analyses. H.M., I.S., T.O., N.Y., Y.Taniguchi,

B.H., H.U., Y.M. contributed to the ALMA observational strategy. N.H.H. independently inspected the ALMA data. G.R.B., T.F., R.P. inspected independently the X-Shooter spectra. G.R.B. contributed to the observations. F.E.B. contributed to the X-Shooter observational strategy. Y.H. and M.O. performed MOSFIRE observations and analysed the archival and our own MOSFIRE data. All authors discussed the results and commented on the manuscript.

**Author Information**   Preprints and permissions information is available at www.nature.com/reprints. The authors declare no competing financial interests. Readers are welcome to comment on the online version of the paper. Correspondence and requests for materials should be addressed to T.H. (thashimoto@est.osaka-sandai.ac.jp).

**Methods**

**Cosmological model.** The adopted cosmological parameters are $H_0 = 70.4 \, \text{km s}^{-1} \, \text{Mpc}^{-1}$, $\Omega_\text{m} = 0.272$, $\Omega_\text{b} = 0.045$, and $\Omega_\Lambda = 0.728$ [25].

**ALMA observations and data reduction.** We observed MACS1149-JD1 with ALMA in Band 7 with a configuration C40-3 (ID 2015.1.00428.S, PI: A. K. Inoue). To cover the uncertainty derived from the photometric redshift analysis[4, 6–9], we used four setups with contiguous frequencies labeled as T3, T4, T5, and T6 encompassing the frequency range $314.4 - 340.5$ GHz and the redshift range $z = 9.0 - 9.8$. In each setup, a total bandwidth of $7.5$ GHz was used, split into four spectral windows (SPWs) each with a $1.875$ GHz bandwidth in the Frequency Division Mode (FDM). Each SPW has a $7.8125$ MHz resolution, corresponding to a velocity resolution of $\sim 7 \, \text{km s}^{-1}$. The total observation times are $75.6$, $35.3$, $119.4$, and $42.3$ minutes, for T3, T4, T5, and T6, respectively. The T3, T4, T5, and T6 data were reduced using the CASA pipeline version 4.7.0, 4.5.2, 4.7.2, and 4.6.0, respectively, with a standard calibration script provided by the ALMA observatory. We then produced final images and cubes with the `CLEAN` task using natural weighting to maximize point-source sensitivity. The spatial resolution is $0''.62 \times 0''.52$ (FWHM) and the beam position angle was PA $= -8.9°$. A quasar, J1229+0203, was used for bandpass and flux calibrations, for which a flux uncertainty is estimated to be $\lesssim 10\%$.

**[O III] 88 $\mu$m line.** To search for a line, we created a data cube, six native channels of which are binned, resulting in a velocity resolution of $\sim 42 \, \text{km s}^{-1}$. In the T5 setup at $\sim 335$ GHz, we found a $> 3.0\sigma$ signal in five continuous binned-channels, where $1\sigma$ is the local noise estimated

with the CASA task `imstat`. This frequency region is free from atmospheric absorption features. We then created a velocity-integrated intensity image between 335.5 and 335.8 GHz. The peak intensity of MACS1149-JD1 is $129.8 \pm 17.5$ mJy km s$^{-1}$ beam$^{-1}$ corresponding to a significance level of $7.4\sigma$, where $1\sigma$ error values in this paper denote the $1\sigma$ rms or standard deviation unless otherwise specified.

The spatial centroid of the emission line is in good positional agreement with that of the UV continuum emission observed by *HST* (Figure 1). Both images are similarly elongated along the gravitational lensing shear. We measured the integrated line flux using the CASA task `imfit` to be $0.229 \pm 0.048$ Jy km s$^{-1}$. To obtain the redshift, we extracted the 1D-spectrum from the region with $> 3\sigma$ signals in the velocity-integrated intensity image. As can be seen, the [O III] line is detected at around 335.6 GHz (or 893.2 $\mu$m) in the Solar system barycentric frame. Applying a Gaussian fit to the line, and with a rest-frame [O III] frequency 3393.006244 GHz, we obtain a redshift $z_{\rm [OIII]}$ = $9.1096 \pm 0.0006$ and FWHM of $154 \pm 39$ km s$^{-1}$, which is reasonable for a low mass galaxy[26]. The integrated flux and redshift leads to an observed luminosity of $(7.4 \pm 1.6) \times (10/\mu) \times 10^7 \, L_\odot$, where $\mu$ is the magnification factor (see later sections for the choice of $\mu$=10).

With the CASA task `imfit`, we obtained the deconvolved size of $(0''.82 \pm 0''.25) \times (0''.30 \pm 0''.14)$. Assuming that lensing effects are equal for the major and minor axes, the intrinsic size is $(3.7 \pm 1.1)/\sqrt{\mu}$ [kpc] $\times$ $(1.4 \pm 0.9)/\sqrt{\mu}$ [kpc].

**Upper limits on dust continuum emission, the total infrared luminosity, and the dust mass.** To create a dust continuum image, we collapsed all four setup data in the frequency range from 314.4 to 340.5 GHz. In this procedure, we excluded the frequency around the [OIII] line, $\pm 3 \times FWHM_{\text{[OIII]}}$. With the CASA task imstat, the rms noise level for the continuum image is 17.7 $\mu$Jy beam$^{-1}$. The panel (a) of Extended Data Figure 1 shows ALMA dust continuum contours of MACS1149-JD1 overlaid on the *HST* F160W image. In the continuum image, we do not detect any signal above $3\sigma$, placing a stringent upper limit on dust continuum emission $S_{\nu,90\mu m} < 5.3 \times (10/\mu)$ $\mu$Jy beam$^{-1}$.

We estimate the total infrared luminosity by integrating the modified-black body radiation over $8 - 1,000$ $\mu$m with the emissivity index of $\beta_{\text{d}} = 1.5$ and the dust temperature of $T_{\text{d}} = 30$–60 K similarly to previous studies[27–30]. The results corrected for the CMB effects[11,31] are listed in Extended Data Table 1. Assuming a dust mass absorption coefficient $\kappa = \kappa_0(\nu/\nu_0)^{\beta_{\text{d}}}$, where $\kappa_0 = 10$ cm$^2$ g$^{-1}$ at 250 $\mu$m[12], we obtain the $3\sigma$ dust mass upper limit of $5.3 \times 10^5 \times (10/\mu)$ $M_\odot$ for the case of 40 K. The dust mass upper limit is increased if the dust temperature is lowered (see Extended Data Table 1). In this estimation, we have assumed the dust emitting region size to be a single beam size following other high-$z$ null detection cases[10,31–33]. If, instead, we assume the same size as that of the [OIII] emitting region, roughly 3 times larger than the beam size, the upper limits can be relaxed by a factor of $\sqrt{3}$ compared with the values listed in Extended Data Table 1.

**VLT/X-Shooter observations and results.** We performed the X-Shooter observations in February and April 2017 (098.A-0534, PI: N. Laporte) for 4.25 and 4 hours, respectively. We adopted a nodding offset of 4".0 so that a bright galaxy entered the slit every other exposure. By detecting the continuum in these half exposures, we confirmed that MACS1149-JD1 was correctly acquired. The chosen slit orientation (+15°) encompasses both the bright galaxy and the elongated shape of the lensed target (Extended Data Figure 2). In order to maximise efficiency, we used 900s, 810s, 740s exposures in the NIR, VIS and UVB arms respectively.

The data were reduced using ESO Reflex tool[34] (v2.8) with the latest version of the X-Shooter pipeline (v 2.9.3). The final spectrum was obtained by combining all reduced exposures using both EsoReflex and a master flat derived from all the data, and the IRAF `imcombine` task on the individual observing blocks. Both methods give similar results. No emission line was found in the UVB and VIS arm spectra. Visual inspection of the NIR arm shows a possible emission line at 12,267.4 Å with an integrated flux of $4.3\pm1.1\times10^{-18}$ erg s$^{-1}$ cm$^{-2}$. Based upon the rms measured in similar sized adjacent apertures, the emission line has a formal significance of $4\sigma$. Taking into account air refraction and the motion of the observatory, we deduce a vacuum wavelength in the barycentric system : $12{,}271.51 \pm 2.27$ Å corresponding to Ly$\alpha$ at a redshift of $9.0942 \pm 0.0019$. We rule out other identifications (e.g. [O II], H$\beta$, [O III] and H$\alpha$) since further emission lines would be seen given our wide wavelength coverage and typical line ratios seen in other star forming sources. Within the likely wavelength range where Ly$\alpha$ might be found, the ratio of resolution elements with positive and negative deviations above a signal/noise ratio of 4 exceeds 5, emphasising that the detection cannot be dismissed as a fluctuation in the noise variation. Although there is some Keck MOSFIRE data for MACS1149-JD1 both within the Keck archive and from a recent campaign, it has insufficient additional exposure time to strengthen the signal/noise of the X-shooter detection.

The redshifts obtained from [OIII] and Ly$\alpha$ are slightly different, corresponding to a velocity offset of $\Delta v_{\rm Ly\alpha} = -450 \pm 60$ km s$^{-1}$ (Extended Data Figure 3). For a single galaxy emitting both [OIII] and Ly$\alpha$, inflowing gas with a high neutral hydrogen column density, $\log N_{\rm HI}/({\rm cm}^{-2}) \simeq 20 - 21$ would lead to Ly$\alpha$ becoming blue-shifted[23,24] and similar offsets have been seen at $z \sim 2-3$[35]. Such inflowing gas would rejuvenate MACS1149-JD1, causing a secondary burst after a relatively inactive phase as indicated by our SED analyses. The Ly$\alpha$ line profile would also be affected by the surrounding intergalactic medium (IGM). Since our SED analysis indicates a dominant phase of earlier star formation, we may expect a large ionised bubble around MACS1149-JD1. Assuming a simple model[20–22], the Strömgren radius can be estimated as $R_{\rm s} = (3N_{\rm ion}/4\pi \langle n_{\rm H} \rangle)^{1/3}$, where $N_{\rm ion}$ is the total ionizing photon number emitted by the early stellar component and $\langle n_{\rm H} \rangle$ is the mean hydrogen density at $z = 9.1$. From the SED analyses, $N_{\rm ion}$ is estimated to be $\simeq 1 \times 10^{70} \times (10/\mu)$ regardless of the star formation history. Assuming an escape fraction of 20%, we obtain a Strömgren radius of $0.4 \times (f_{\rm esc}/0.2)^{1/3}(10/\mu)^{1/3}$ Mpc, corresponding to a velocity offset $\simeq 500 \times (f_{\rm esc}/0.2)^{1/3}(10/\mu)^{1/3}$ km s$^{-1}$, where $f_{\rm esc}$ corresponds to the escape fraction of ionizing photons from galaxies to the intergalactic medium.

Alternatively, it is possible that MACS1149-JD1 comprises two $z = 9.1$ galaxies separated kinematically by 450 km s$^{-1}$. In this case, the IRAC excess could be associated with either galaxy but, regardless, the Ly$\alpha$ emitting component would lie within the ionized bubble. Indeed, we note a $z = 7.1$ galaxy has Ly$\alpha$, [OIII], and [CII] each offset spatially and in velocity space[17]. This galaxy shows a $\sim 500$ km s$^{-1}$ blue-shifted Ly$\alpha$ with respect to [OIII].

**Astrometry.** Several studies have demonstrated that there are, in some cases, spatial offsets between ALMA-detected sources and their *HST*-counterparts[29,36]. We took advantage of a dusty spiral galaxy, CLASH 2882[37–39], in our ALMA field-of-view to examine possible astrometric off-sets. The panel (b) of Extended Data Figure 1 shows ALMA dust continuum contours of CLASH 2882 overlaid on the *HST* F160W image. Dust continuum is clearly detected at the exact position of the NIR counterpart of CLASH 2882 at the significance level of $\simeq 10\sigma$ (where $1\sigma = 16.0$ $\mu$Jy beam$^{-1}$). The centroids of ALMA and *HST* are consistent within $0''.1$.

As can be seen from the *HST* morphology shown on Figure 1 and the panel (a) of Extended Data Figure 1 MACS1149-JD1 is elongated and may be formed by two components at $z = 9.1$ separated by $0''.33$. To determine the exact position of the Ly$\alpha$ emitting region, we collapsed the final 2D spectrum in wavelength direction, and measured the separation between Ly$\alpha$ and the continuum of the reference galaxy. Within the positional uncertainty $\simeq 0''.7$, the Ly$\alpha$ emitting region is consistent with arising in either component. As revealed in Figures 2 and 3, the strong excess in IRAC 4.5$\mu$m is a prominent feature of MACS1149-JD1. To determine from which component the IRAC 4.5$\mu$m emission arises, we first checked the astrometry between *HST* and IRAC 4.5$\mu$m with bright point source stars. Within the $1\sigma$ uncertainty $\simeq 0''.3$ and the resolution of the IRAC camera, the centroid of the 4.5$\mu$m channel is consistent with emission from either of the two *HST* counterparts.

**Magnification factor.** With our new spectroscopic redshift, we revisited the magnification calculations using all the mass models released in the framework of the Frontier Fields survey [40–46]. The magnification values range from $\mu=4.9^{+0.4}_{-0.5}$ (Diego v4.1) to $89.0^{+192.7}_{-40.8}$ (CATS v4.1) with a mean value of $\langle\mu\rangle=21.8^{+260.0}_{-17.4}$. However, if we remove the most extreme values, the mean magnification factor decreases to $\langle\mu\rangle=11.7^{+80.1}_{-5.1}$, where the error bar takes into account the dispersion of values from all the Frontier Fields models. For simplicity we assume a fiducial value for the magnification factor of 10 in this study.

**Spectral energy distribution modeling.** We built the SED of our target by combining data from the 3 deep *HST* images obtained within the framework of the Frontier Fields survey[7], the deep K$_s$ image obtained with VLT/HAWK-I[16] as well as the deep *Spitzer*/IRAC 3.6 and 4.5 $\mu$m data[7]. In addition, we use our deep upper limit on dust continuum emission as well as our measurement of the [O III] flux described above.

We use our own SED fitting code[47]. Briefly, the stellar population synthesis model of GALAXEV[48] including nebular continuum and emission lines[49] is used. The [O III] line flux is estimated from metallicity and SFR using semi-empirical models[10, 19]. The ionizing photon escape fraction is fixed to zero. A Calzetti's law[50] is assumed for dust attenuation, and empirical dust emission templates[51] are adopted. The Chabrier initial mass function[52] at $0.1 - 100\ M_\odot$ is adopted, and a mean IGM model is applied[53]. The model parameters and their steps are summarized in Extended Data Table 2. To estimate the best-fit parameters, we use the least $\chi^2$ formula[54] including an analytic treatment of upper limits for the non-detection. Uncertainties on the parameters are estimated based on a Monte Carlo technique.

In a first attempt, we adopted a single exponentially declining star formation history (SFH). The panel (a) of Extended Data Figure 4 shows the best-fit SED obtained with a reduced $\chi^2_\nu \simeq 6.9$ and a stellar age of $\simeq 300$Myr. Although the stellar age is comparable to that obtained by previous studies[7, 16], this model could not reproduce the [O III] line flux observed with ALMA. The situation is worse if we assume a constant SFH which has $\chi^2_\nu \simeq 11.4$ leading to a stellar age of $\simeq 500$ Myr (the panel (b) of Extended Data Figure 4).

We next considered a two stellar component SED comprising a young starburst and an old population. In order to reproduce the observed SED with as small a number of parameters as possible, we assumed constant SFRs for both components. Extended Data Figure 5 gives a schematic picture of our models. If one assumes an exponentially declining (rising) SFH for the old component, there is always an equivalent constant SFH model with a lower (higher) age of the population. We adopted several star formation durations for the old component, namely $\tau = 10, 100, 200, 300$, and 400 Myr. In these two component models, we can reproduce the observations with $\chi^2_\nu \sim 2.8 - 3.0$. Models with $\tau = 300$ and 400 Myr lead to a stellar age $> 400$ Myr at $z = 9.1$, similar to the age of the Universe at this redshift, which we consider physically unacceptable. The other three models give a similar quality fit and ages for the old stellar population ranging from 230 to 320 Myr. Due to the similarity of the physical and statistical results produced by the three remaining models, we cannot easily discriminate one over the other. For this reason, we select the $\tau = 100$ Myr model as our fiducial case but include uncertainties which reflect the standard deviation caused by the choice of a $\tau = 10$ or 200 Myr model (Extended Data Table 3). The panels (c) and (d) of Extended Data Figure 4 show the two best-fit SEDs for $\tau = 10$ and 200 Myr.

Although we can exclude contamination of the IRAC channel 2 from the [OIII] 5007 Å line at the [OIII] redshift $z = 9.1096$ (and the lower Ly$\alpha$ redshift), conceivably there is contamination from intense H$\beta$ and [OIII] 4959 Å[13]. Our nebular emission line model[49] already accounts for such contributions up to a rest-frame equivalent width (H$\beta$+[OIII] 4959 Å) of $\sim 1550$ Å in young metal-poor cases. Nevertheless, Extended Data Figure 6 show that such young metal-poor models cannot reproduce the red IRAC colour because of the contribution of strong nebular continuum and [OII] 3727 Å line emission in the IRAC channel 1. Such an explanation for the IRAC excess is further disfavoured by e.g, the observed dust emission upper limit (panel (c) of Extended Data Figure 6).

**Dynamical mass.** We derive the dynamical mass, $M_{\rm dyn}$, of MACS1149-JD1 following analyses at $z \sim 2$ [55]. Because MACS1149-JD1 does not show a velocity gradient in [OIII], we assume that MACS1149-JD1 is dispersion-dominated. In this case, the dynamical mass can be obtained as $M_{\rm dyn} = 6.7\sigma_{\rm line}^2 r_{\frac{1}{2}}/G$, where $\sigma_{\rm line}$ is the line velocity dispersion, $r_{\frac{1}{2}}$ is the half light radius, and $G$ is the gravitational constant. Taking the lensing effect into account, the intrinsic half light radius becomes $(1.9 \pm 0.6)/\sqrt{\mu}$ with an uncertainty of a factor $1/\sqrt{\mu}$ depending on the direction of elongation. We thus obtain $M_{\rm dyn} = (4 \pm 3) \times (10/\mu)^{0.5} \times 10^9 \, M_\odot$. Therefore, the dynamical mass is comparable to or larger than the stellar mass, $(1.1 \pm 0.5) \times (10/\mu) \times 10^9 \, M_\odot$.

**Predictions for future *JWST* observations.** The star formation history deduced from the SED-fits allows us to trace the earlier luminosity evolution of MACS1149-JD1 as a function of cosmic age. Accordingly, we computed the SED from $z = 15$ to 9.1 and find that the source would have been as bright as $m_{1500}$=26 AB at $z = 12$ (corrected for magnification; Extended Data Figure 6). Such a source would be easily detectable by the *JWST*. Between $z \sim 12$ and $z \leq 15$, it would be detected with NIRCam at $10\sigma$ within 20 minutes. Furthermore and according to NIRSpec sensitivity the UV continuum and a significant Ly$\alpha$ break of MACS1149-JD1 analogues will be detectable at a $5\sigma$ level in less than an hour (Extended Data Figure 7).

**Code availability.** The ALMA data were reduced using the CASA pipeline versions 4.7.0, 4.5.2, 4.7.2, and 4.6.0, for the T3, T4, T5, and T6 data, respectively. Our SED fitting code is our custom one, which is available at https://www.astr.tohoku.ac.jp/∼mawatari/KENSFIT/KENSFIT.html with instructions.

**Data availability.** This paper makes use of the following ALMA data: ADS/JAO.ALMA#2015.1.00428.S, available at https://almascience.nao.ac.jp/aq/?project_code=2015.1.00428.S. This paper is also based on observations made with ESO Telescopes at the La Silla Paranal Observatory under programme ID 098.A-0534, available at http://archive.eso.org/wdb/wdb/eso/sched_rep_arc/query?progid=098.A-0534(A). *HST* and *Spitzer* as well as VLT/HAWK-I photometry data that support our finding are available in the IOPscience repository with the identifiers DOI: 10.3847/1538-4357/aa5d55[7] and DOI: 10.3847/1538-4357/aaa9c2[16], respectively. The datasets generated during and/or analysed during the current study are available from the corresponding author on reasonable request.

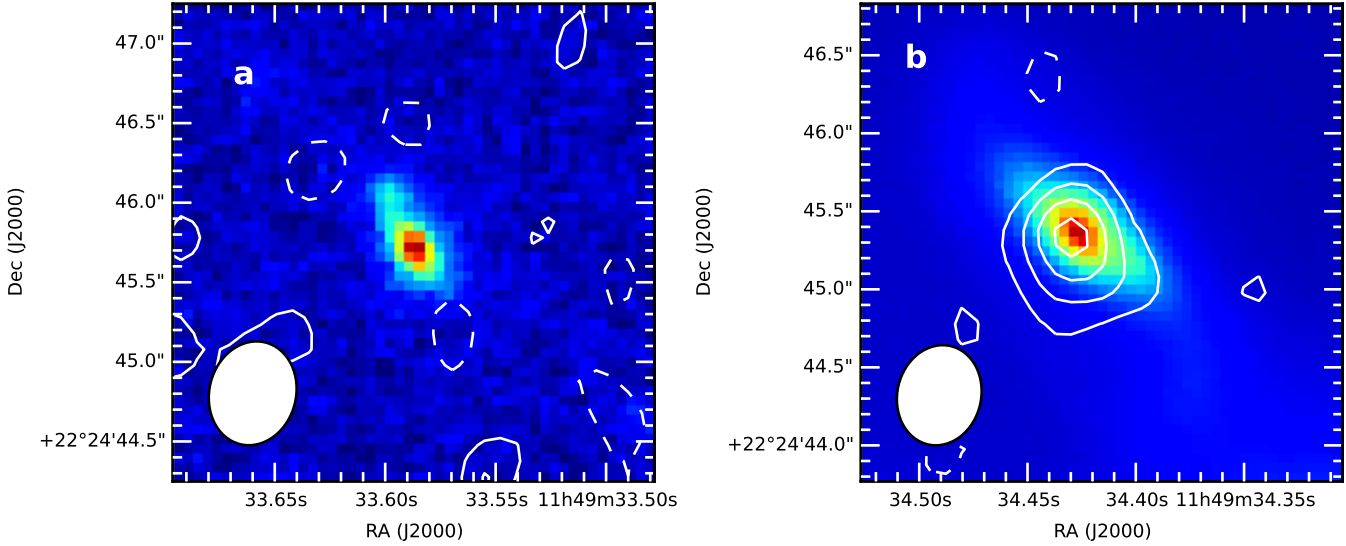

**Extended Data Figure 1 | ALMA dust contours of MACS1149-JD1 and a serendipitous continuum object.** (a) ALMA dust contours of MACS1149-JD1 overlaid on the *HST* F160W image. Contours are drawn at $\pm 2\sigma$, where $\sigma = 17.7$ $\mu$Jy beam$^{-1}$. Negative contours are shown by the dashed line. Ellipse at the lower left corner indicates the synthesized beam size of ALMA. (b) Dust continuum of a dusty galaxy at $z = 0.99$ in our ALMA field-of-view, overlaid on the *HST* F160W image. Contours are drawn at $(-2, 2, 4, 6, 8, 9.5) \times \sigma$, where $1\sigma = 16.0$ $\mu$Jy beam$^{-1}$.

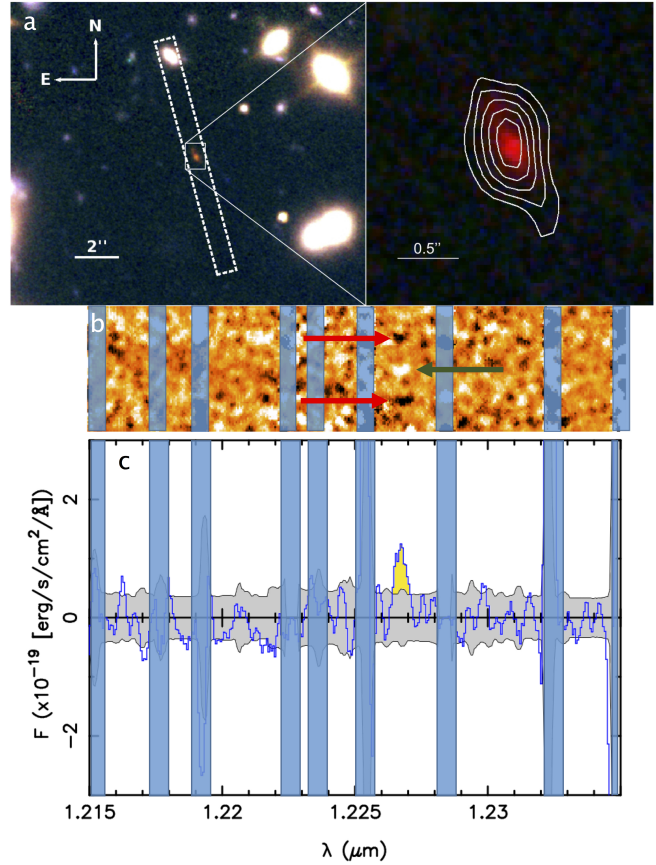

**Extended Data Figure 2 | X-Shooter observations and the Lyα spectra of MACS1149-JD1.** (a) Orientation of the X-Shooter slit (white dashed rectangle) demonstrating the successful acquisition of MACS1149-JD1 via the alignment of the slit to follow its lensed elongation as well as the inclusion of a bright foreground galaxy. (b) X-Shooter 2D spectra of MACS1149-JD1 with the position of Lyα marked with a green arrow, and the two negative counterparts with red arrows. Sky lines are highlighted by blue rectangles. (c) 1D extracted spectra in a 0".8 aperture. Lyα is indicated in yellow, $2\sigma$ is in grey and the sky lines are marked by blue rectangles.

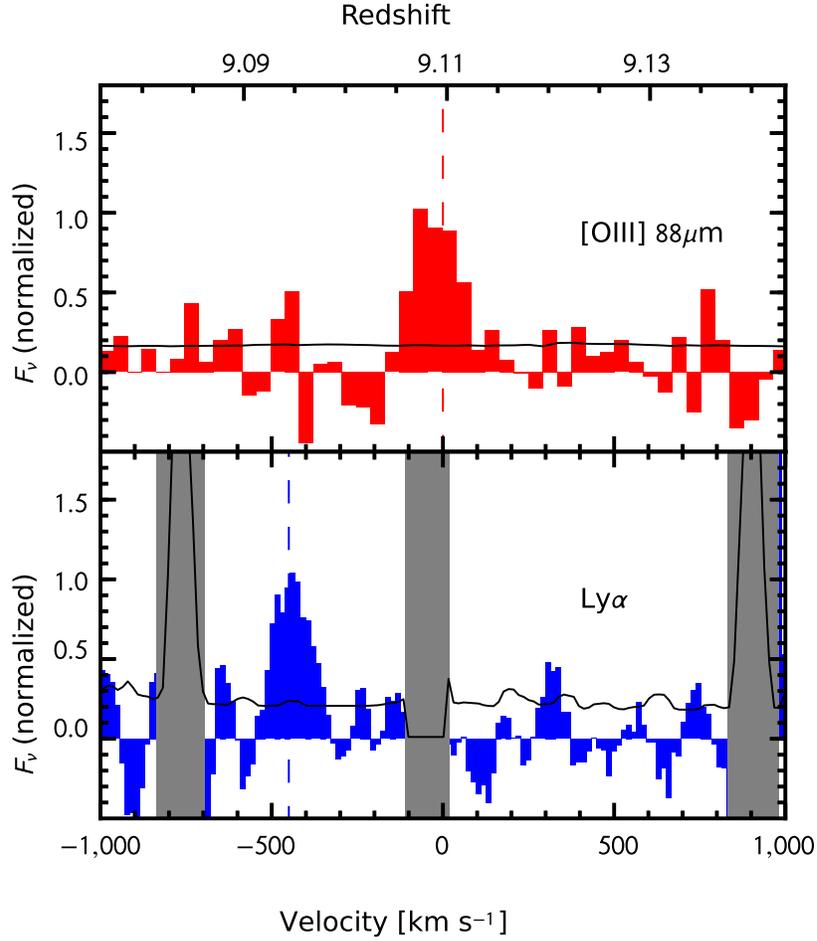

**Extended Data Figure 3 | ALMA [OIII] 88 $\mu$m and X-Shooter Ly$\alpha$ spectra in velocity space.** [OIII] and Ly$\alpha$ are shown with a resolution of $\sim 42$ km s$^{-1}$ and 15 km s$^{-1}$, respectively. The values are normalized by the peak flux densities. The velocity zero point corresponds to the [OIII] redshift $z = 9.1096$ (red dashed line) and the Ly$\alpha$ offset is $\simeq 450$ km s$^{-1}$ (blue dashed line). Grey rectangles show regions contaminated by night sky emission. The data at around $-100$ to $0$ km s$^{-1}$ is removed from the analysis because the night sky emission is too strong. The black solid lines indicate the $1\sigma$ noise level for the velocity resolutions.

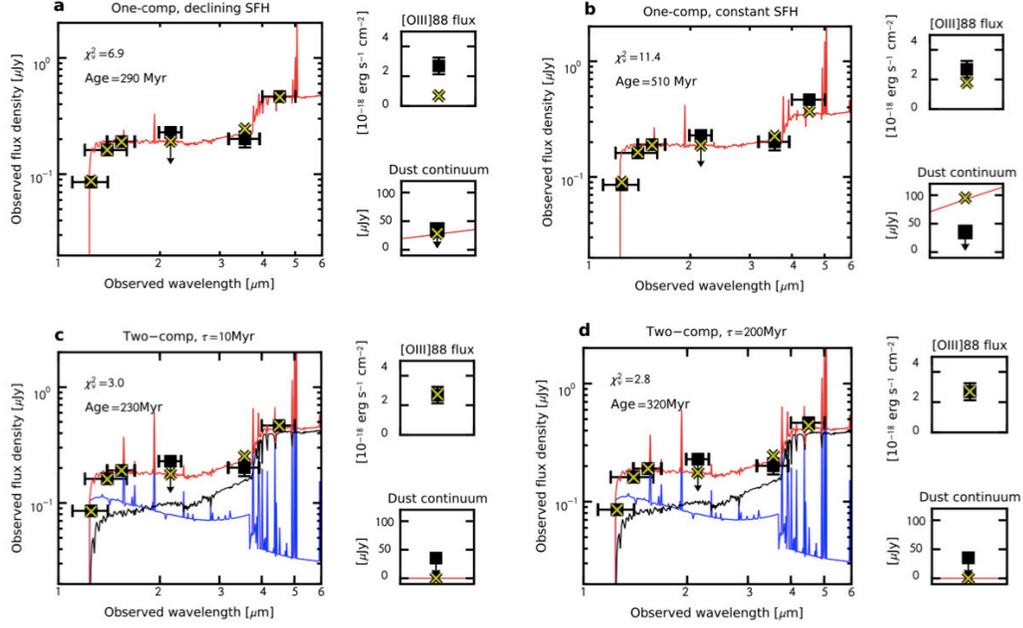

**Extended Data Figure 4 | Best-fit SEDs of MACS1149-JD1 with various star formation histories.** Panels (a) and (b) show the best-fit SED with a single stellar component assuming an exponentially declining and constant SFHs, respectively. Panels (c) and (d) show the best-fit SEDs with two stellar components assuming a constant SFH. The star formation durations of the old component are $\tau = 10$ and 200 Myr. The reduced $\chi^2$ value, $\chi^2_\nu$, and the best-fit stellar age for each model is shown in the upper left corner. The meanings of the symbols are the same as those in Figure 2.

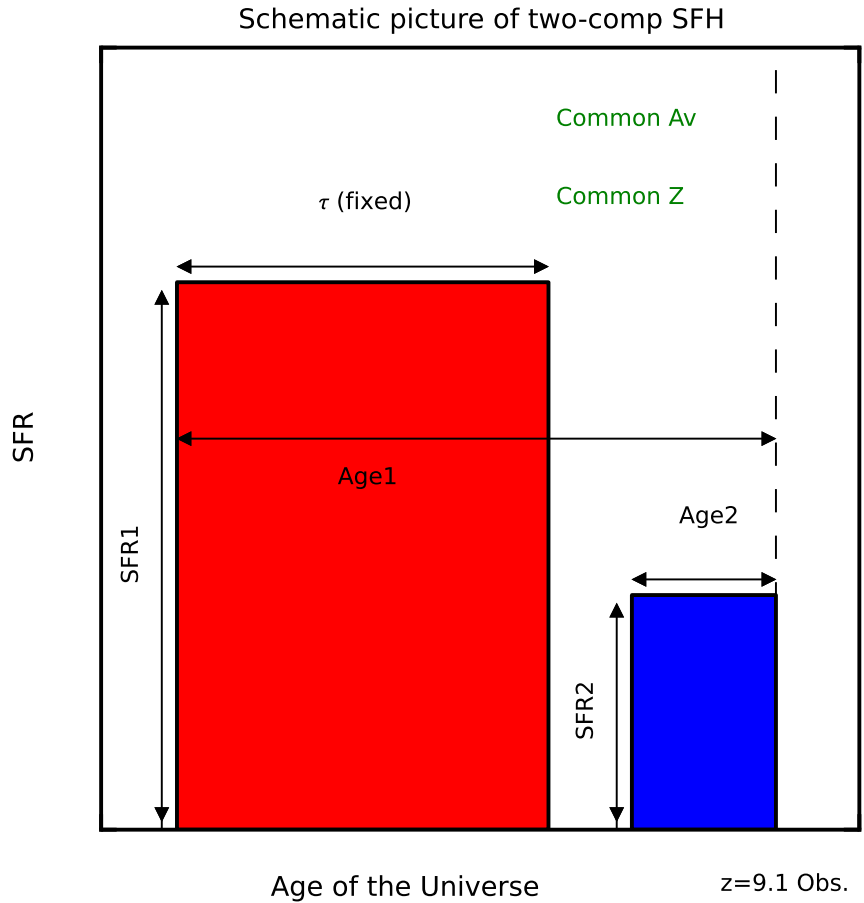

**Extended Data Figure 5 | Schematic picture of star formation histories of our two components models.** The Red and blue rectangles show the old and young stellar components with constant SFRs, respectively. The old components stops its star formation activity after a fixed duration of $\tau$. The black vertical dashed line indicates the observation at $z = 9.1$ (the Universe age $\simeq 550$ Myr). Each component is described with Age and SFR parameters. For simplicity, both components have a common dust attenuation $A_{\rm v}$ and metallicity $Z$.

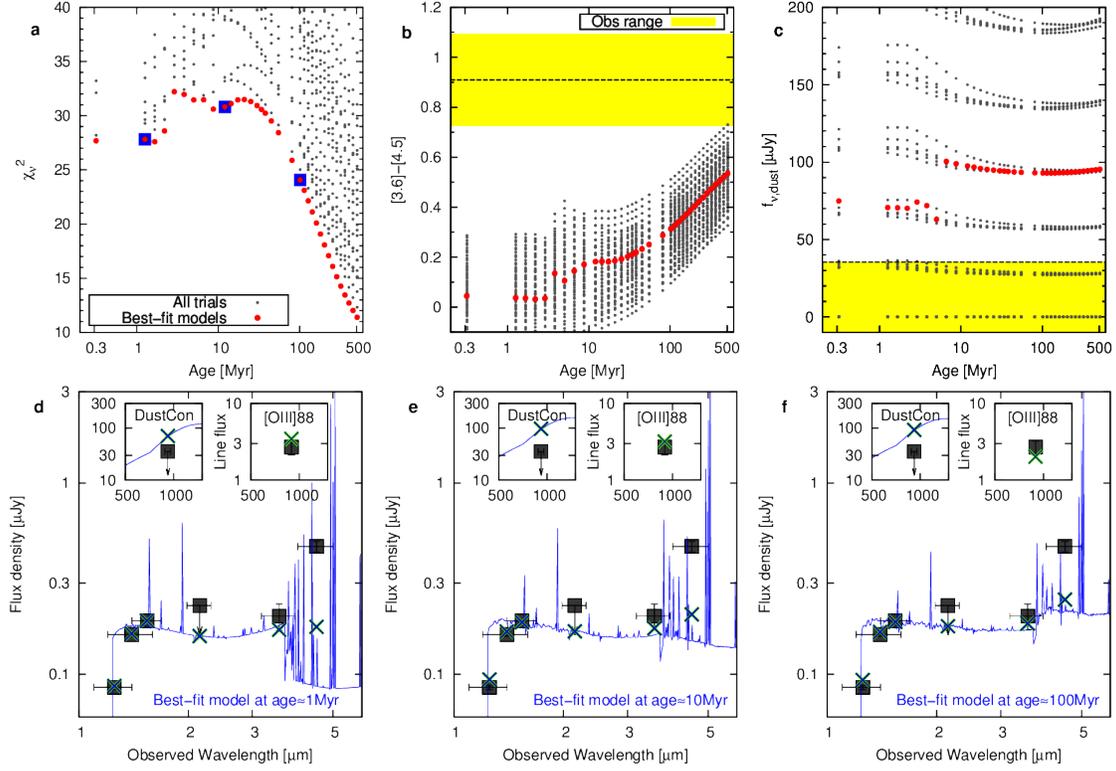

**Extended Data Figure 6 | Comparisons of constant SFH models and observational constraints.** We show the (a) $\chi_\nu^2$, (b) IRAC colour, and (c) dust emission, plotted against stellar age. All model grids are shown with small grey dots, while the best-fit models at given stellar ages are indicated with red circles. In panel (b), the black horizontal dashed line indicates the observed value and the yellow shaded region its $1\sigma$ uncertainty. In panel (c) the black horizontal dashed line refers to the $2\sigma$ upper limit. Panels (d) - (f) show the best-fit SEDs at ages of 1, 10 and 100 Myr indicated by the blue squares in panel (a). Inset figures show the dust continuum flux density ($\mu$Jy) and [OIII] 88 $\mu$m flux ($10^{-18}$ erg cm$^{-2}$ s$^{-1}$). Panel (d) demonstrates that a strong nebular continuum plus [OII] 3727 Å emission counteracts intense $H\beta$ plus [OIII] 4959 Å emission producing an 'inverse Balmer break' for very young metal-poor cases.

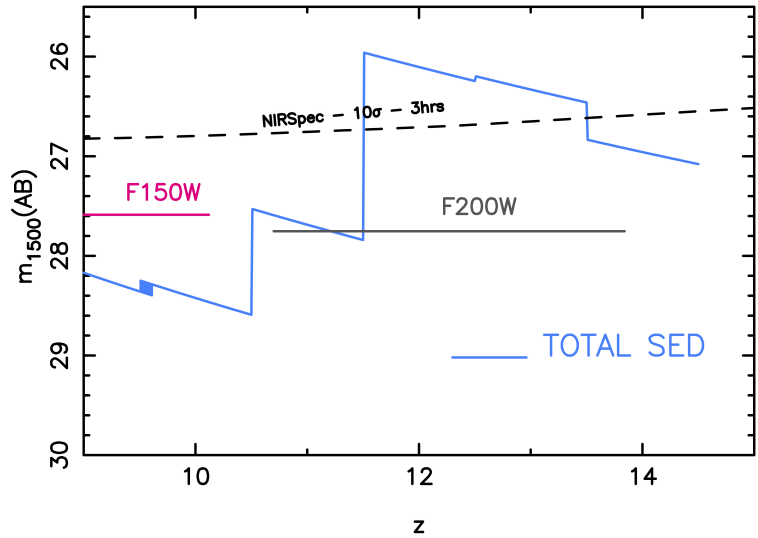

**Extended Data Figure 7 | Evolution of the UV luminosity of MACS1149-JD1 as a function of redshift.** For each redshift bin ($\Delta z=1$), we extrapolated the magnitude by assuming a constant Star Formation Rate over the redshift interval (blue curve). We over-plotted the sensitivity of NIRCAM filters (pink and grey) covering the $\lambda 1500$Å rest-frame ($10\sigma$ in $\approx 20$ minutes) and the NIRSpec sensitivity (dashed black line) at the same wavelength ($10\sigma$ in 3 hours).

| $T_{\rm d}$ | $\beta_{\rm d}$ | $L_{\rm TIR}$ | $M_{\rm d}$ |
|---|---|---|---|
| (K) | | ($10^9\ L_\odot$) | ($10^5\ M_\odot$) |
| 30 | 1.5 | $< 6.0\times (10/\mu)$ | $< 20.0\times (10/\mu)$ |
| 40 | 1.5 | $< 7.7\times (10/\mu)$ | $< 5.3\times (10/\mu)$ |
| 50 | 1.5 | $< 11.4\times (10/\mu)$ | $< 2.3\times (10/\mu)$ |
| 60 | 1.5 | $< 17.3\times (10/\mu)$ | $< 1.3\times (10/\mu)$ |

**Extended Data Table 1** | **Upper limits on the infrared luminosity and dust mass.**

Upper limits represent the $3\sigma$ limits. The total luminosity, $L_{\rm TIR}$, is estimated by integrating the modified-black body radiation at $8 - 1,000$ $\mu$m. We use dust temperatures ranging from $T_{\rm d} = 30$ K to $60$ K and the emissivity index $\beta_{\rm d} = 1.5$. The dust mass, $M_{\rm d}$, is estimated with a dust mass absorption coefficient $\kappa = \kappa_0(\nu/\nu_0)^{\beta_{\rm d}}$, where $\kappa_0 = 10$ cm$^2$ g$^{-1}$ at $250$ $\mu$m (Ref.[12]).

| | |
|---|---|
| **Single stellar component** | |
| Star formation history | Exponential ($\tau = \pm 0.03, 0.06, 0.1, 0.3, 0.6, 1.0$ and $10$ Gyr) |
| Age [Myr] | $0.1 - 550$ (age of the Universe) |
| Dust $A_V$ [mag] | $0 - 1$ |
| Metallicity | $Z = 0.0001, 0.0004, 0.004, 0.008$, and $0.02$ |
| Star formation rate [$M_\odot$ yr$^{-1}$] | arbitrary (amplitude) |
| **Two stellar components** | **(Constant + shut down)** |
| Age1 [Myr] | $0.1 - 550$ (age of the Universe) |
| Age2 [Myr] | $0.1 - 550$ (age of the Universe) |
| Common Dust $A_V 1 = A_V 2$ [mag] | $0 - 1$ |
| Common Metallicity $Z1 = Z2$ | $Z = 0.0001, 0.0004, 0.004, 0.008$, and $0.02$ |
| Total Star formation rate [$M_\odot$ yr$^{-1}$] | arbitrary (amplitude) |
| SFR ratio $f_{\rm SFR0} \equiv ({\rm SFR2/SFR1})_{z=9.1}$ | $10^{-4} - 10^4$ |

**Extended Data Table 2 | Summary of our SED parameters.** In single component models, three SFHs of constant, exponentially-declining, and exponentially-rising SFR are considered. In two components models, a constant SFH with star formation durations of $\tau = 10, 100, 200, 300$, and $400$ are used for the old component (see Extended Data Figure 4 for a schematic picture of our models). The parameter $f_{\rm SFR0}$ defines the SFR ratio of the two components at the observation ($z = 9.1$). In all models, dust attenuation steps are $\Delta A_V = 0.1$, and age steps follow the original $36$ steps in GALEXEV[48] between $0.1 - 550$ Myr. The SFR ratio steps are $\Delta \log(f_{\rm SFR0}) = 0.1$. The metallicity value is in the absolute unit with a Solar metallicity of $0.02$.

| Parameters | Values |
| --- | --- |
| $\chi^2$ | 5.52 |
| $\nu$ | 2 |
| $\chi^2_\nu$ | 2.76 |
| Age$_{\rm old}$ [Myr] | $290^{+190}_{-120}$ |
| Age$_{\rm young}$ [Myr] | $3^{+2}_{-1}$ |
| Metallicity | $0.004^{+0.004}_{-0.004}$ |
| $A_V$ [mag] | $0.0^{+0.1}_{-0.0}$ |
| Stellar mass [$10^9 M_\odot$] | $1.1^{+0.5}_{-0.2} \times (10/\mu)$ |
| Star formation rate [$M_\odot$ yr$^{-1}$] | $4.2^{+0.8}_{-1.1} \times (10/\mu)$ |
| $f_{SFR0}$ | $4^{+76}_{-3}$ |
| $N_{\rm ion,max}$ [$10^{70}$] | $\simeq 1 \times (10/\mu)$ |
| $R_{\rm s,max}$ [Mpc] | $\simeq 0.7 \times (10/\mu)^{1/3}$ |
| Velocity shift$_{\rm max}$ [km s$^{-1}$] | $\simeq 900 \times (10/\mu)^{1/3}$ |
| $t_{\rm inact}$ [Myr] | $180^{+20}_{-60}$ |
| $z_{\rm f,old}$ | $15.4^{+2.7}_{-2.0}$ |

**Extended Data Table 3 | Summary of SED fits in our fiducial model.** Summary of the best-fit results with our fiducial SED model. The uncertainties reflect any standard deviation caused by the choice of a $\tau = 10$ or $200$ Myr model. The best-fit SED is shown in Figure 2, and its SFH is shown in red in Figure 3. $N_{\rm ion,max}$, $R_{\rm s,max}$, and Velocity shift$_{\rm max}$ denote the maximum numbers of ionizing photons, Strömgren radius, and the shift of the Ly$\alpha$ damping wing, respectively. $t_{\rm inact}$ is a duration of the inactive phase, and $z_{\rm f,old}$ is the formation redshift of the old component.

**References**

1. Robertson, B., Ellis, R., Furlanetto & S., Dunlop, J. Cosmic Reionization and Early Star-forming Galaxies: A Joint Analysis of New Constraints from Planck and the Hubble Space Telescope. *Astrophys. J. Lett.* **802L**, 19 (2015)

2. Bouwens, R. et al. Reionization After Planck: The Derived Growth of the Cosmic Ionizing Emissivity Now Matches the Growth of the Galaxy UV Luminosity Density. *Astrophys. J.* **811**, 140 (2015)

3. Stark, D. Galaxies in the First Billion Years After the Big Bang. *Annual Review of Astronomy and Astrophysics.* **54**, 761-803 (2016)

4. McLeod, D., McLure, R. & Dunlop, J. The $z = 9 - 10$ galaxy population in the Hubble Frontier Fields and CLASH surveys: the $z = 9$ luminosity function and further evidence for a smooth decline in ultraviolet luminosity density at $z \geq 8$. *Mon. Not. R. Astron. Soc.* **459**, 3812-3824 (2016).

5. Oesch, P. et al. The Dearth of $z \sim 10$ Galaxies in all HST Legacy Fields – The Rapid Evolution of the Galaxy Population in the First 500 Myr *arXiv* https://arxiv.org/abs/1710.11131

6. Zheng, W. et al. A magnified young galaxy from about 500 million years after the Big Bang. *Nature* **489**, 406-408 (2012).

7. Zheng, W. et al. Young Galaxy Candidates in the Hubble Frontier Fields. IV. MACS J1149.5+2223. *Astrophys. J.* **836**, 210, doi:10.3847/2041-8213/aa794f (2017).

8. Huang, K. et al. Spitzer UltRa Faint SUrvey Program (SURFS UP). II. IRAC-Detected Lyman-Break Galaxies at $6 < z < 10$ Behind Strong-Lensing Clusters. *Astrophys. J. Lett.* **817**, 11 (2016).

9. Kawamata, R. et al. Precise Strong Lensing Mass Modeling of Four Hubble Frontier Field Clusters and a Sample of Magnified High-redshift Galaxies. *Astrophys. J.* **819**, 114 (2016).

10. Inoue A. K. et al. Detection of an oxygen emission line from a high-redshift galaxy in the reionization epoch. *Science* **352**, 1559-1562 (2016).

11. da Cunha, E. et al. On the Effect of the Cosmic Microwave Background in High-redshift (Sub-)millimeter Observations. *Astrophys. J.* **766**, 13 (2013).

12. Hildebrand, R. H. The Determination of Cloud Masses and Dust Characteristics from Submillimetre Thermal Emission. *Quarterly J. R. Astron. Soc.* **24**, 267 (1983).

13. Smit, R. et al. High-precision Photometric Redshifts from Spitzer/IRAC: Extreme [3.6] - [4.5] Colors Identify Galaxies in the Redshift Range $z \sim 6.6 - 6.9$. *Astrophys. J.* **801**, 122 (2015).

14. Roberts-Borsani, G. et al. $z \geq 7$ Galaxies with Red Spitzer/IRAC [3.6]-[4.5] Colors in the Full CANDELS Data Set: The Brightest-Known Galaxies at $z \sim 7 - 9$ and a Probable Spectroscopic Confirmation at $z = 7.48$. *Astrophys. J.* **823**, 143 (2016).

15. Stark, D. et al. Ly$\alpha$ and C III] emission in $z = 7 - 9$ Galaxies: accelerated reionization around luminous star-forming systems? *Mon. Not. R. Astron. Soc.* **464**, 469-479 (2017).

16. Hoag, A. et al. HST Grism Observations of a Gravitationally Lensed Redshift $9.5$ Galaxy. *Astrophys. J.* **854**, 39, doi:10.3847/1538-4357/aaa9c2 (2018).

17. Carniani, S. et al. Extended ionised and clumpy gas in a normal galaxy at $z = 7.1$ revealed by ALMA. *Astron. Astrophys.* **605**, 42 (2017).

18. Vernet, J. et al. X-shooter, the new wide band intermediate resolution spectrograph at the ESO Very Large Telescope. *Astrophys. J.* **536**, A105 (2011).

19. Inoue, A. K. et al. ALMA Will Determine the Spectroscopic Redshift $z > 8$ with FIR [OIII] Emission Lines. *Astrophys. J. Lett.* **780**, L18 (2014).

20. Cen, R. & Haiman, Z. Quasar Strömgren Spheres Before Cosmological Reionization. *Astrophys. J.* **542**, L75-L78 (2000).

21. Haiman, Z. The Detectability of High-Redshift Ly$\alpha$ Emission Lines prior to the Reionization of the Universe. *Astrophys. J.* **576**, L1-L4 (2002).

22. Hu, E. et al. An Ultraluminous Ly$\alpha$ Emitter with a Blue Wing at $z = 6.6$. *Astrophys. J. Lett.* **825**, L7 (2016).

23. Dijkstra, M., Haiman, Z. & Spaans, M. Ly$\alpha$ Radiation from Collapsing Protogalaxies. II. Observational Evidence for Gas Infall. *Astrophys. J.* **649**, 14-36 (2006).

24. Verhamme, A. et al. Using Lyman$-\alpha$ to detect galaxies that leak Lyman continuum. *Astron. Astrophys.*, **578**, A7 (2015).

25. Komatsu, E. et al. Seven-year Wilkinson Microwave Anisotropy Probe (WMAP) Observations: Cosmological Interpretation. *Astrophys. J.* **192**, 18 (2011).

26. Hashimoto, T. et al. A Close Comparison between Observed and Modeled Ly$\alpha$ Lines for $z \sim 2.2$ Ly$\alpha$ Emitters. *Astrophys. J.* **812**, 157 (2015).

27. Watson, D. et al. A dusty, normal galaxy in the epoch of reionization. *Nature* **519**, 327-330 (2015).

28. Knudsen, K. K. et al. A merger in the dusty, $z = 7.5$ galaxy A1689-zD1? *Mon. Not. R. Astron. Soc.* **466**, 138-146 (2017).

29. Laporte, N. et al. Dust in the Reionization Era: ALMA Observations of a $z = 8.38$ Gravitationally Lensed Galaxy. *Astrophys. J.* **837**, L21 (2017).

30. Faisst, A. K. et al. Are High-redshift Galaxies Hot? Temperature of $z > 5$ Galaxies and Implications for Their Dust Properties. *Astrophys. J.* **847**, 21 (2017).

31. Ota, K et al. ALMA Observation of 158 $\mu$m [C II] Line and Dust Continuum of a $z = 7$ Normally Star-forming Galaxy in the Epoch of Reionization. *Astrophys. J.* **792**, 34 (2014).

32. Ouchi, M. et al. An Intensely Star-forming Galaxy at $z \sim 7$ with Low Dust and Metal Content Revealed by Deep ALMA and HST Observations. *Astrophys. J.* **778**, 102 (2013).

33. Schaerer, D. et al. New constraints on dust emission and UV attenuation of $z = 6.5 - 7.5$ galaxies from millimeter observations. *Astron. Astrophys.* **574**, A19 (2015).

34. Freudling, W. et al. Automated data reduction workflows for astronomy. The ESO Reflex environment. *Astrophys. J.* **559**, A96 (2013).

35. Trainor, R., Steidel, C., Strom, A., & Rudie, G. The Spectroscopic Properties of Ly$\alpha$-Emitters at $z \sim 2.7$: Escaping Gas and Photons from Faint Galaxies. *Astrophys. J.* **809**, 89 (2015).

36. González-López, J. et al. The ALMA Frontier Fields Survey. I. 1.1 mm continuum detections in Abell 2744, MACS J0416.1−2403 and MACS J1149.5+2223. *Astron. Astrophys.* **597**, A41 (2017).

37. Dwek, E. et al. Dust Formation, Evolution, and Obscuration Effects in the Very High-redshift Universe. *Astrophys. J.* **788**, L30 (2014).

38. Dwek, E. et al. Submillimeter Observations of CLASH 2882 and the Evolution of Dust in this Galaxy. *Astrophys. J.* **813**, 119 (2015).

39. Zavala, J. A. et al. Early science with the Large Millimeter Telescope: dust constraints in a $z \sim 9.6$ galaxy. *Mon. Not. R. Astron. Soc.* **453**, L88-L92 (2015).

40. Richard, J. et al. Mass and magnification maps for the Hubble Space Telescope Frontier Fields clusters: implications for high-redshift studies. *Mon. Not. R. Astron. Soc.* **444**, 268-289 (2014).

41. Johnson, T. L. et al. Lens Models and Magnification Maps of the Six Hubble Frontier Fields Clusters. *Astrophys. J.* **797**, 48 (2014).

42. Ishigaki, M. et al. Hubble Frontier Fields First Complete Cluster Data: Faint Galaxies at $z \sim 5 - 10$ for UV Luminosity Functions and Cosmic Reionization. *Astrophys. J.* **799**, 12 (2015).

43. Keeton, C. R. On modeling galaxy-scale strong lens systems. *General Relativity and Gravitation* **42**, 2151-2176 (2010).

44. Liesenborgs, J., De Rijcke, S. & Dejonghe, H. A genetic algorithm for the non-parametric inversion of strong lensing systems. *Mon. Not. R. Astron. Soc.* **367**, 1209-1216 (2006).

45. Diego, J. M., Protopapas, P., Sandvik, H. B. & Tegmark, M. Non-parametric inversion of strong lensing systems. *Mon. Not. R. Astron. Soc.* **360**, 477-491 (2005).

46. Merten, J. et al. Creation of cosmic structure in the complex galaxy cluster merger Abell 2744. *Mon. Not. R. Astron. Soc.* **417**, 333-347 (2011).

47. Mawatari, K. et al. Possible identification of massive and evolved galaxies at $z \simeq 5$. *Publ. Astron. Soc. Jpn.* **68**, 46 (2016).

48. Bruzual, G. & Charlot, S. Stellar population synthesis at the resolution of 2003. *Mon. Not. R. Astron. Soc.* **344**, 1000-1028 (2003).

49. Inoue, A. K. Rest-frame ultraviolet-to-optical spectral characteristics of extremely metal-poor and metal-free galaxies. *Mon. Not. R. Astron. Soc.* **415**, 2920-2931 (2011).

50. Calzetti, D. et al. The Dust Content and Opacity of Actively Star-forming Galaxies. *Astrophys. J.* **533**, 682-695 (2000).

51. Rieke, H. et al. Determining Star Formation Rates for Infrared Galaxies. *Astrophys. J.* **692**, 556-573 (2009).

52. Chabrier, G. Galactic Stellar and Substellar Initial Mass Function. *Pub. Astron. Soc. Pac.* **115**, 763-795 (2003).

53. Inoue, A. K. et al. An updated analytic model for attenuation by the intergalactic medium. *Mon. Not. R. Astron. Soc.* **442**, 1805-1820 (2014).

54. Sawicki, M. SEDfit: Software for Spectral Energy Distribution Fitting of Photometric Data. *Pub. Astron. Soc. Pac.* **124**, 1208 (2012).

55. Förster Schreiber, M. et al. The SINS Survey: SINFONI Integral Field Spectroscopy of $z \sim 2$ Star-forming Galaxies. *Astrophys. J.* **706**, 1364-1428 (2009).